\documentclass{jpp}
\usepackage{graphicx}
\usepackage{todonotes}
\usepackage{mathtools}

\usepackage[utf8]{inputenc}
\usepackage[T1]{fontenc}
\usepackage{amsmath}
\usepackage{xcolor}
\usepackage[hidelinks]{hyperref}

\shorttitle{Particle Distribution in Rotating and Tandem Traps}
\shortauthor{G.X. Li, E.J. Kolmes, I.E. Ochs and N.J. Fisch}

\title{Approximating the Particle Distribution in Rotating and Tandem Mirror Traps}

\author{G. X. Li\aff{1}      
    \corresp{\email{greta.li@princeton.edu}}, E. J. Kolmes\aff{2}, I. E. Ochs\aff{2} \and N. J. Fisch\aff{2}}

\affiliation{\aff{1}Department of Physics, Princeton University,  Princeton, NJ 08544, USA
\aff{2}Department of Astrophysical Sciences, Princeton University, Princeton, NJ 08544, USA}

\begin{document}

\maketitle

\begin{abstract}
Steady state distribution functions can be used to calculate stability conditions for modes, radiation energy losses, and particle loss rates. Heuristic analytic approximations to these distributions can capture key behaviors of the true distributions such as the relative speeds of different transport processes while possessing computational advantages over their numerical counterparts. In this paper, we motivate and present a closed-form analytic model for a distribution of particles in a centrifugal or tandem mirror. We find that our model outperforms other known models in approximating numerical steady-state simulations outside of a narrow range of low confining potentials. We demonstrate the model's suitability in the high confining potential regime for applications such as loss cone stability thresholds, fusion yields, and available energy. 
\end{abstract}

\section{Introduction} \label{sec:introduction}
There has been growing interest in open field line configurations for magnetic plasma confinement. This includes both conventional mirrors, centrifugal mirrors, and tandem mirrors. In centrifugal traps, rotation is induced by an $\vec{E}\times\vec{B}$ drift, which introduces a centrifugal potential that improves plasma confinement \citep{lehnert_rotating_1971, bekhtenev_problems_1980, ellis_experiment_2001, teodorescu_confinement_2010} and can improve plasma stability \citep{bekhtenev_problems_1980, abdrashitov_hot_1991, white_centrifugal_2018, kolmes_loss-cone_2024}. For tandem mirrors, end plugs generate a confining potential in the center cell. Magnetic mirrors possess practical advantages over other confinement configurations that are important to consider in the path to commercial fusion. In particular, the open-ended axial magnetic system of mirrors is relatively simpler to build and maintain than the closed toroidal systems of tokamaks and stellarators, and mirror systems can contain high beta plasmas in steady-state operation \citep{post_magnetic_1987}. 

Advancements in mirror technology as demonstrated in GAMMA-10 \citep{cho_observation_2005}, the Maryland Centrifugal Mirror eXperiment (MCMX) \citep{ellis_steady_2005}, and the gas dynamic trap (GDT) \citep{ivanov_gas-dynamic_2013, ivanov_gdt_2017} have paved the path for further mirror development with experiments such as the Wisconsin HTS Axisymmetric Mirror (WHAM) \citep{egedal_fusion_2022}, the Centrifugal Mirror Fusion eXperiment (CMFX) \citep{white_centrifugal_2018}, the SMOLA device \citep{Sudnikov2017}, and multiple mirror (MM) traps \citep{beery_plasma_2018,miller_rate_2021}. 

Apart from their use in fusion, centrifugal mirror confinement is effective in differential containment devices, namely devices used for mass separation \citep{weibelprl1980,gueroultclean2018, gueroulthaz2015,gueroultpsst2014,gueroultpop2014,ohkawapop2002,fettermanpop2011,oilerppr2024, hidekumaprl1974, hiroenf1975, dolgolenko_separation_2017,timofeevphysuspekhi2014,voronapan2015, litvakArchimedesPlasmaMass2003}. The regime of large plasma rotation in centrifugal mirror devices is where differentiating particles on the basis of mass is particularly effective. It also happens to be the regime of the greatest pertinence here.

In the literature on standard (non-centrifugal) mirrors, studies have heavily relied on heuristic approximations to the steady-state particle distribution in phase space \citep{rosenbluth_highfrequency_1965,post_electrostatic_1966,Tang1972,Smith1984,Smith1984ii, Futch1972, newcomb_equilibrium_1981}. This approach has been particularly productive for calculations of stability conditions for loss-cone modes. However, despite the important role played by kinetic physics in the centrifugal or tandem traps, the corresponding approximations are far less well-developed in the literature on mirror traps with an arbitrary confining potential.

Previous approaches to capturing rotating loss-cone behavior include applying prefactors that vanish at the loss-cone boundary \citep{volosov_characteristics_1969,turikov_effect_1973,khudik_longitudinal_1997}, modifying a Maxwellian distribution to suppress values in the loss cone \citep{catto_particle_1985, kolmes_loss-cone_2024,kolmes_upper_2024}, and leveraging other known distributions with some loss-cone features and thermal anisotropy \citep{summers_kappa_2025}. 
 
The purpose of this paper is to consider how best to construct a heuristic approximation for the distribution of particles in a low collisionality rotating or tandem mirror. We propose a new analytic closed-form model that improves upon past models and reflects the relative dominance of pitch angle scattering compared to energy scattering. We also examine how different heuristic approximations perform in different contexts and how best to understand the suitability of a given approximation for a given application. 

The paper is structured as follows. \hyperref[sec: prior]{Section 2} reviews previous models which motivate the general form of the class of solutions that we consider in \hyperref[sec: func]{Section 3}. \hyperref[sec: code]{Section 4} covers the numerical scheme from \citet{Ochs2023AmbipolarPotentials} that we use to validate our model. \hyperref[sec: pref]{Section 5} presents our analytic approximation explicitly and a useful parametrization of space. \hyperref[sec: gof]{Section 6} analyzes error of our model compared to numerical simulations, finding that our model outperforms other known models in the regime where the confining potential $\phi \gtrsim 2$. \hyperref[sec: apps]{Section 7} examines the suitability of the model in three applications: loss cone stability thresholds, fusion yields, and free energy. To conclude, \hyperref[sec: final]{Section 8} discusses the suitability and potential adjustments of our model for different confinement regimes. 

\section{Prior Models} \label{sec: prior}
The distribution of particles in a centrifugal trap is typically not Maxwellian. Rather, these traps have loss cones: regions of phase space in which particles are not confined. It is constructive to explicitly review parameters and equations relevant to rotating plasmas in magnetic mirrors. We then examine previous proposed models before reducing our search to functions of a certain form.

Suppose we have a mirror machine with magnetic field strength $B$, which is greatest at the ends and smallest at the mid-plane, and confining potential $\varphi$, which could include the electrostatic potential and centrifugal potential due to rotation \citep{post_magnetic_1987}. We define the mirror ratio $R_0$ as 
\begin{equation}
    R_0 \coloneqq \frac{B_{\text{end}}}{B_{\text{mid}}},
\end{equation}
where $B_\text{end}$ and $B_{\text{mid}}$ are the magnetic fields evaluated at the mirror ends and mid-plane respectively.
We also define a dimensionless difference of confining potential evaluated at the ends and the mid-plane, 
\begin{equation}
    \phi \coloneqq \frac{\varphi_{\text{end}}-\varphi_{\text{mid}}}{T}.
\end{equation}
Here $\varphi_{\text{end}}$ and $\varphi_{\text{mid}}$ are the respective potentials at the ends and at mid-plane of the device, including both centrifugal and electrostatic components, and $T$ is the plasma temperature.

Assuming that the first adiabatic invariant, the magnetic moment, is conserved, we can derive the confinement condition from \citet{post_magnetic_1987}, 
\begin{equation}
    0 \leq \phi + R_0 x^2 \sin^2(\theta) - x^2, \label{eqn:lossCone}
\end{equation}
where $x$ is the dimensionless speed of the particle (normalized to the thermal speed) and $\theta$ is the pitch angle, which is the angle between the direction of the velocity and the direction of the magnetic field.

The loss cone region of momentum-space contains all the points $(x, \theta)$ such that the above inequality does not hold. Any particles that diffuse into the loss cone region are lost from the system. Increasing either the mirror ratio $R_0$ or the generic confining potential $\phi$ will shrink the loss cone region, thus improving the confinement of the system.

We briefly review prior proposed models. One such analytic model is the truncated Maxwellian, which is the Maxwellian distribution function with values in the loss cone sent to zero as used in \citet{kolmes_loss-cone_2024},
\begin{equation}
    f_{\text{tm}}(x, \theta) = A \left[\pi^{-3/2} e^{-x^2} \right] \Theta\left(\phi + R_0 x^2 \sin^2(\theta)-x^2 \right),
\end{equation}
where $A$ is the normalization constant such that the integral of momentum-space distribution over all space evaluates to unity and $\Theta$ is the Heaviside step function that cuts out the loss cone region. 

This model has the advantage in its relative simplicity and recovery of the Maxwellian distribution in the limit as $\phi \rightarrow \infty$. The loss cone vertex is located at $(\sqrt{\phi}, 0)$, so in this limit, the loss cone region shifts far away from the bulk of the particle distribution. However, $f_{\text{tm}}$ has distinctly un-physical behavior at the loss cone boundary due to the infinite gradient there. This becomes an issue for applications like fusion yield calculations in cases where the confining potential is not large. 

For the steady-state distribution $f$ solution to the linearized Fokker-Planck equation, \citet{najmabadi_collisional_1984} find an analytic prefactor $g \coloneqq f/f_M$, where $f_M$ is the Maxwellian and $g$ has adjustable parameters,
\begin{equation}
    g(x, \theta) = 1 - q_0 \ln\left(\frac{e^{a^2}+e^{x^2} + \sqrt{\rho^2 + (e^{a^2} + e^{x^2})^2}}{e^{a^2}-e^{x^2} + \sqrt{\rho^2 + (e^{a^2} - e^{x^2})^2}}\right),
\end{equation}
where
\begin{equation}
    \begin{aligned}
    q_0 & \coloneqq \left(\ln\left(\frac{w+1}{w-1}\right)\right)^{-1} \quad w \coloneqq \left(1 + \frac{1}{Z_{\perp, s} R_0}\right)^{1/2} \\
    a & \coloneqq \left(\ln(w) + \phi\right)^{1/2} \qquad \rho \coloneqq \left(\frac{2x^2}{Z_{\perp, s}}\right)^{1/2} e^{x^2} \tan(\theta).
    \end{aligned}
\end{equation}
The species-dependent constant $Z_\perp$ is defined as the following for ion species $a$ and electron species $e$,  
\begin{subequations}\label{eqn:zperpnaj}
\begin{gather}
Z_{\perp, a} \coloneqq 
\frac{1}{2} \left(\frac{\sum\limits_{b} n_b Z_b^2 \lambda_{ab}}{\sum\limits_{b} n_b Z_b^2 \lambda_{ab} m_a T_b/ m_b T_a}\right) \\
Z_{\perp, e} \coloneqq \frac{1}{2} \left(1 + \frac{\sum\limits_{b} n_b Z_b^2 \lambda_{eb}}{n_e \lambda_{ee}} \right)
\end{gather}
\end{subequations}
where $Z_b$, $m_b$, $n_b$, $T_b$ are the charge, mass, density, and temperature of other species $b$, respectively. $\lambda_{ab}$ is the Coulomb logarithm. Faster species are excluded from the sums. 

We removed the correction term $1/4 x^2$ in $\rho$ and $w$ to avoid divergence at the origin. This formulation does not go to zero at the loss cone boundary, but for the sake of later comparison, we apply a Heaviside step function $\Theta$ that cuts out the loss cone region and then re-normalize the distribution. One advantage of this model is the explicit dependence on $Z_\perp$. However, in order to calculate the collisional end loss rates, the focus is on approximating the region within one e-folding of the loss cone vertex. We will see later that this leads to a weaker approximation for the bulk of the distribution if the vertex is sufficiently far away from the origin.

An earlier analytic model used by \citet{volosov_characteristics_1969} and \citet{turikov_effect_1973} is as follows, 
\begin{equation}
    f_{\text{vol}}(x, \theta) =  A \left[\sqrt{\phi + R_0 x^2 \sin^2(\theta) - x^2} e^{-x^2 \sin^2 \left(\theta \right)} \right] \Theta\left(\phi + R_0 x^2 \sin^2(\theta) - x^2\right),
\end{equation}
where $A$ is the normalization constant. When $\phi = 0$  and $R_0 = 2$, $f_{\text{vol}}$ reduces to the analytic model of a standard non-rotating mirror in \citet{rosenbluth_highfrequency_1965}. The prefactor term in the square root ensures that the model smoothly decays to zero at the loss cone. However, in the limit as $\phi \rightarrow \infty$, the analytic model does not recover the Maxwellian distribution due to the exponential term's dependence on only the perpendicular component of momentum, $x \sin(\theta)$.

More recently, subtracted-kappa distributions have been suggested to approximate loss-cone-like regions for space plasmas by \citet{summers_kappa_2025}, particularly in contexts where anisotropy is important. However, we will restrict our scope here to models that vanish within the loss cone described by Eq. \eqref{eqn:lossCone}.

Trial functions $h \coloneqq f/f_M$ have also been used to find particle loss rates in mirrors, where $h$ serves as a prefactor to the Maxwellian distribution $f_M$ \citep{catto_collisional_1981, catto_particle_1985, khudik_longitudinal_1997}. In particular, an iso-contours approach is used where lines of constant $h$ can be used in their integral calculations. However, although their application did not require or provide an explicit analytic expression for $h$, we find this approach informative and will pursue it further.

\section{Function Properties}\label{sec: func}
It is convenient to write candidate models as the product of a truncated Maxwellian and some prefactor function.
Note, only the restrictions on the prefactor itself limit the class of allowable functions for the distribution function. Let our model be $f_{\text{mod}}$ and prefactor be $g_{\text{mod}}$, 
\begin{equation} \label{eq:fmod}
    f_{\text{mod}}(x, \theta; R_0, \phi) = A \left[g_{\text{mod}}(x, \theta; R_0, \phi) \left(\pi^{-3/2} e^{-x^2}\right) \right]\Theta\left(\phi + R_0 x^2 \sin^2(\theta) - x^2 \right)
\end{equation}
where $R_0$ and $\phi$ are included as parameters to emphasize the model's dependency on them. We provide the explicit normalization constant, 
\begin{gather}
    A = \left( \int dx d\theta \hspace{3pt} 2 \pi x^2 \sin(\theta) \Big[g_{\text{mod}} f_{\text{tm}} \Big] \right)^{-1}
\end{gather}
where the $2 \pi$ comes from the distribution's cylindrical symmetry in momentum-space after assuming gyrotropy and $f_{\text{tm}}$ is the truncated Maxwellian model. Our prefactor $g_{\text{mod}}$ is of the following form,
\begin{equation}
    g_{\text{mod}}(x, \theta; R_0, \phi) = \frac{f_{\text{mod}}(x, \theta; R_0, \phi)}{f_{\text{tm}}(x, \theta; R_0, \phi)}
\end{equation}
up to the normalization constant.
For example, the normalized truncated Maxwellian would correspond to the constant prefactor, 
\begin{equation}
    g_{\text{tm}}(x, \theta; R_0, \phi)=1.
\end{equation}

Instead of directly analytically approximating $f_{\text{mod}}$ to some numerical steady-state distribution $f_{\text{sim}}$, we can instead fit the prefactor $g_{\text{mod}}$ to a weighted simulation, which is defined as $g_{\text{sim}} \coloneqq f_{\text{sim}}/f_{\text{tm}}$. This quantity measures how similar the simulation value is to the truncated Maxwellian value at a point. We will require our prefactor to force a smooth decay to zero at the loss cone while having negligible effect on the truncated Maxwellian in the limit as $\phi \rightarrow \infty$.

Furthermore, our mirror system has cylindrical symmetry due to the lack of azimuthal angle dependency. This symmetry forces us to consider a prefactor such that 
\begin{equation}\label{eqn:sym}
    \frac{\partial g_\text{mod}}{\partial \theta} \Big\vert_{\theta = \pi/2} = 0,
\end{equation}
where $\theta = \pi/2$ corresponds to the plane of momentum phase space for particles with velocities perpendicular to the magnetic field.

\section{Simulation Set Up} \label{sec: code}
\begin{figure}
\centering
\includegraphics[width=0.98\textwidth]{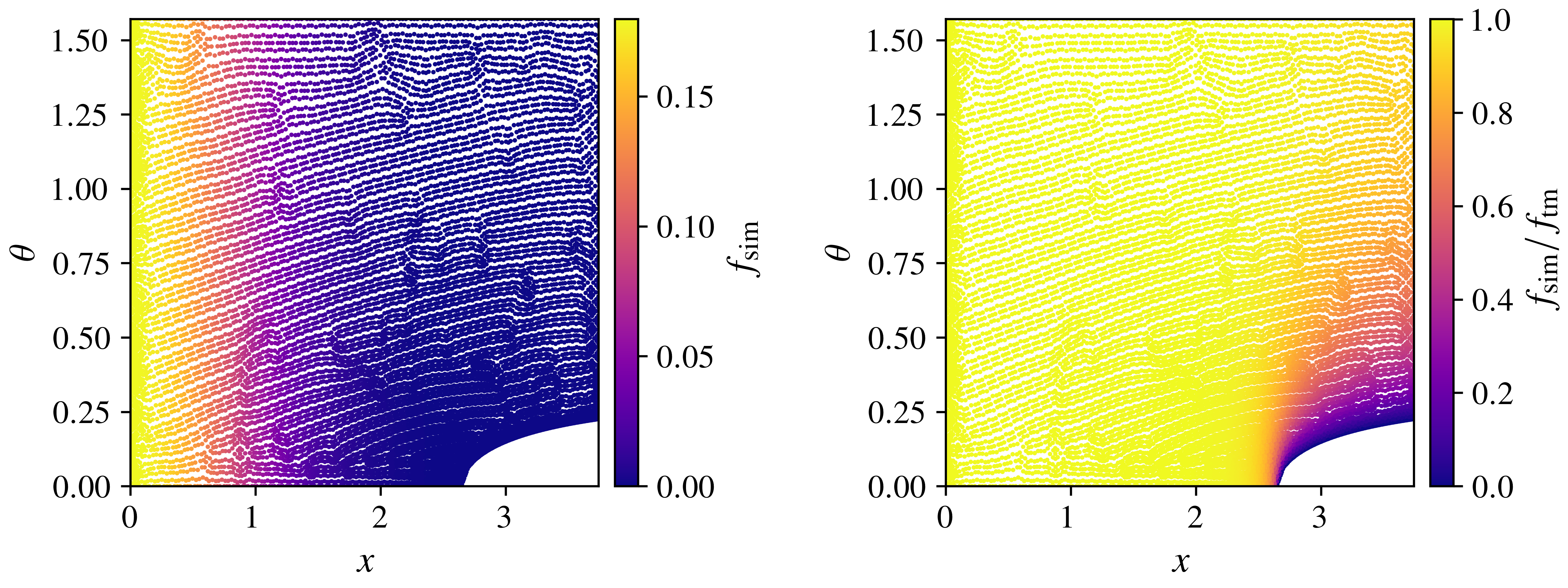}
\caption{On the left, we numerically simulate $f_{\text{sim}}(x, \theta)$ for $\phi=7$,  $R_0=10$, $Z_\perp = 1$, and $K = 7$. The simulation is close to Maxwellian as evidenced by its near independence of $\theta$ and exponential decay in $x$. On the right, we plot the weighted simulation $g_{\text{sim}} \coloneqq f_{\text{sim}}/f_{\text{tm}}$ for the same parameters.} 
\label{fig:simandgsim}
\end{figure}

In order to validate different features of these heuristic models, we use Fokker-Planck simulations. Here, we review those simulations.
Namely, we explain relevant parameters and coordinates as well as choice of source, temperature, and species. Furthermore, we expand on desired properties for the prefactor $g_{\text{mod}}$ that are tailored to our numerical simulation's domain. 

To obtain a numerical steady-state particle distribution $f_{\text{sim}}$ of a magnetic mirror, we simulate the collisional process governed by the Fokker-Planck diffusion equation for an arbitrary species $a$ \citep{Ochs2023AmbipolarPotentials, najmabadi_collisional_1984, rosen2025},
\begin{equation}
    \tau_{0, a} \frac{\partial f_a}{\partial t} = \frac{1}{x^2} \frac{\partial}{\partial x} \left(Z_{\parallel, a} f_a + \frac{1}{2x}\frac{\partial f_a}{\partial x} \right) + \frac{1}{x^3} \left(Z_{\perp, a} - \frac{1}{4x^2} \right) \frac{\partial}{\partial \xi} \Bigg[\left( 1-\xi^2\right) \frac{\partial f_a}{\partial \xi}\Bigg]
\end{equation}
where we have followed the convention in broader mirror literature by assuming gyrotropy and using the following dimension-less coordinates, 
 \begin{equation}\label{eqn:coord}
    x = \frac{v}{v_{th, a}} \qquad \xi = \frac{v_\parallel}{v} \qquad v_{th, a} \coloneqq \sqrt{2T_a/m_a}. 
 \end{equation}
Here $v_{th, a}$ is the thermal speed, $v_{||}$ is the component of the velocity parallel to the magnetic field, $m_a$ is the particle mass, and $\tau_{0, a}$ is the collision time of species $a$. Intra-species collisions are included.

The coordinate-transformed diffusion equation contains two species-dependent quantities, $Z_{\parallel, a}$ and $Z_{\perp, a}$, 
\begin{equation}
    Z_{\parallel, a} \coloneqq \frac{\sum\limits_{b} n_b Z_b^2 \lambda_{ab}/m_b}{\sum\limits_{b} n_b Z_b^2 \lambda_{ab} T_b/m_b T_a} \qquad Z_{\perp, a} \coloneqq \frac{1}{2} \frac{\sum\limits_{b} n_b Z_b^2 \lambda_{ab}}{\sum\limits_{b} n_b Z_b^2 \lambda_{ab} m_a T_b/ m_b T_a}
\end{equation}
where $Z_b$, $m_b$, $n_b$, $T_b$ are the charge, mass, density, and temperature of other species $b$, respectively. $\lambda_{ab}$ is the Coulomb logarithm. $Z_{\perp, a}$ was defined above in Eqn. \eqref{eqn:zperpnaj} as the constant for an ion species. Since the electron species expression, $Z_{\perp, e}$, is consistent with the ion species expression after ignoring ion-electron collisions, we redefine $Z_{\perp, a}$ to hold for either ion or electron species. To repeat, faster species are excluded from the sums. We take $Z_\parallel = 1$, which corresponds to species having the same temperature. As an example of $Z_\perp$ values, a pure, single-ion species plasma has a $Z_\perp$ value of $0.5$. 

We include appropriate modifications to the diffusion equation such as changing the denominator of operators to avoid divergent behavior and adding a source term. Zero flux is enforced at the loss cone boundary as we consider low-collisionality mirror machines. We simulate $K$ e-foldings of the distribution past the loss cone vertex. That is, the simulation domain boundary is $x_\text{max} = \sqrt{\phi + K}$. The simulation implements the finite-element method using the DolfinX library. More details of the numerical simulation setup can be found in \citet{Ochs2023AmbipolarPotentials}. 

Furthermore, we choose to neglect relativistic effects. For the rest of the paper, we find it more convenient to work with the pair of coordinates $(x, \theta)$, where $x$ is the dimensionless speed defined in Eqn. \eqref{eqn:coord} and $\theta$ is pitch angle,
\begin{equation}
 \theta \coloneqq \cos^{-1}(\xi).
\end{equation}

 We place a cold Maxwellian source $f_s(x, \theta; T_s)$ at the origin of momentum-space, 
 \begin{equation}
    f_{s}(x, \theta; T_s) = \pi^{-3/2} e^{-x^2/T_s}.
\end{equation}
 $T_s$ is the parameter determining the relative strength of the source's physical temperature to the steady-state system's background distribution physical temperature. The collision operator assumes a fixed hotter Maxwellian background distribution that the source particles collide with. Cold sources correspond to smaller values of $T_s$, which reduces the source's effect on the overall steady-state particle distribution. For the rest of the paper, we typically choose $T_s = 0.1$ as a sufficiently small value to be considered cold. 

To understand simulation-specific constraints on $g_{\text{mod}}$, we look at the weighted simulation $g_{\text{sim}}$. An example of $f_{\text{sim}}$ and $g_{\text{sim}}$ is shown in Fig. \ref{fig:simandgsim}. Our simulations are limited to positive $x$ and $\theta \in [0, \pi/2]$. We can find our prefactor expression for this region, and our constraint of cylindrical symmetry in Eqn. \eqref{eqn:sym} allows us to extend the expression to all of momentum space.

Note, in this momentum-space Fokker-Planck equation, the relative velocity of colliding particles is approximated with the velocity of the hotter, non-thermal particle. Due to this approximation and the effect of the source term, the simulated solution will somewhat worsen for $x<1$, impacting the low $\phi$ regime then. Our model is intended as a general approximation that focuses on the tail behaviors of distributions with a Maxwellian bulk, agnostic of any source terms. In the low $\phi$ regime, the distribution is more sensitive to the specific source term, so finding any single general model is inherently challenging.

\section{Finding Our Prefactor} \label{sec: pref}
We present a simple logarithmic prefactor to the truncated Maxwellian. To do this, we observe a useful parametrization of momentum-space that reduces our problem to two dimensions. We then fit a logarithmic expression. To improve our model near the loss cone vertex, we implement a `shift' to our parametrization and evaluate its effectiveness.  

\begin{figure}
\centering
\includegraphics[width=0.98\textwidth]{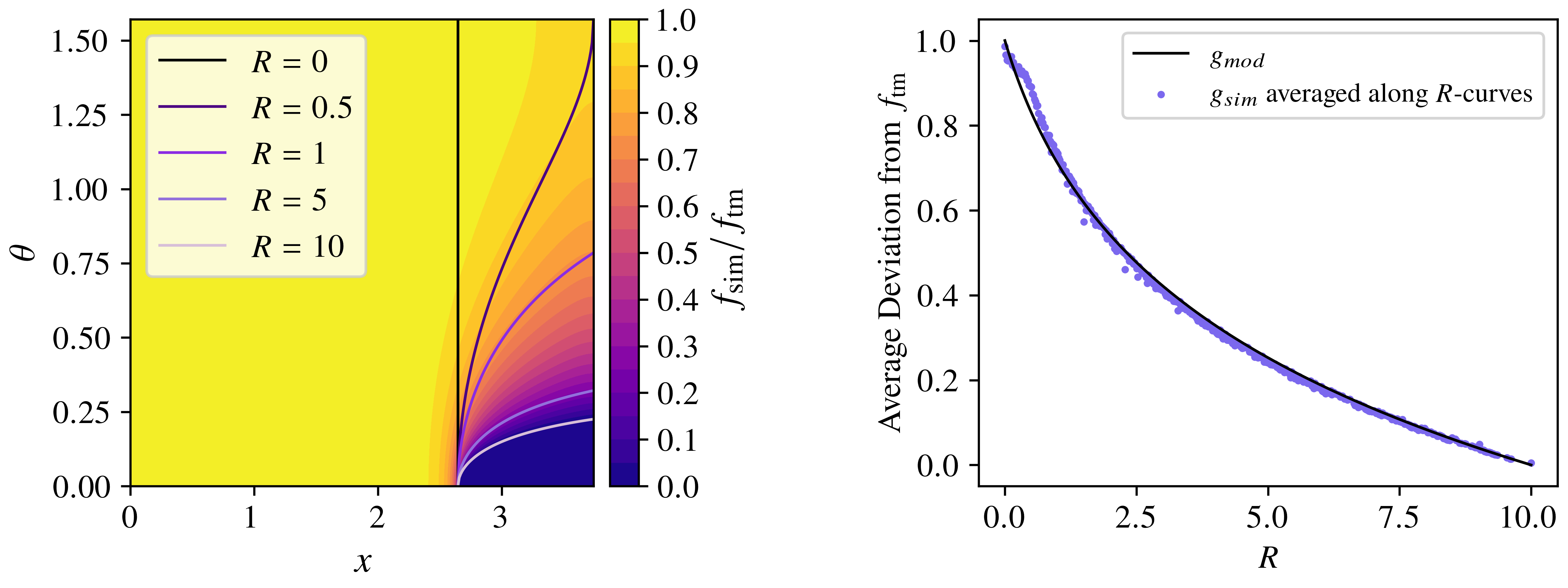}
\caption{To the left, we have the $g_{\text{sim}}$ contour plot for $\phi=7$ and $R_0 = 10$. We have added some loss cone curves used for parametrization. To the right, we plot $g_{\text{sim}}$ with respect to $R$ for the same $\phi$ and $R_0$. For each $R=R_c$ value, we averaged all the $g_{\text{sim}}$ values from points that had $R$ values within $0.002$ of $R_c$. Note the close fit to the overlaid proposed prefactor for high $R$ values, which corresponds to the region around the physical loss cone.} 
\label{fig:curveanddev}
\end{figure}

\subsection{Parametrization by Constant R Curves}
It is convenient to simplify our problem from finding a function of two variables to one. If we can identify a parameter $p$ that associates a value to each level curve of $g_{\text{sim}}$ and also some independent parameter $\eta$, we can change our momentum-space coordinates from $(x, \theta)$ to $(p, \eta)$. Our prefactor $g_{\text{mod}}$ should only have dependence on parameter $p$ since we want $g_{sim}$ and $g_{\text{mod}}$ to have the same level curves. 

Our intuition for a possible parametrization lies in the loss cone condition. Setting equality and rearranging, the loss cone boundary can be written as the following, 
\begin{equation}
R_0 = \frac{x^2 - \phi}{x^2 \sin^2(\theta)}.
\end{equation}
    
Let us vary $R_0$ without changing $\phi$. This results in another loss cone curve that goes through the same vertex as our physical loss cone. As shown in Fig. \ref{fig:curveanddev}, we notice that there is good agreement with these effective loss cone curves with the level curves of the weighted simulation. This motivates us to choose our independent parameter $R$ defined as the following, 
\begin{equation}\label{eq:unshiftedR}
R(x, \theta) = \frac{x^2 - \phi}{x^2 \sin^2(\theta)}, 
\end{equation}
where constant-$R$ curves in momentum-space closely match the level curves of the weighted simulation. 

In other words, there is a physical value of $R(x,\theta)$ that corresponds to the loss cone for the actual, physical magnetic field in the system, but it turns out that the loss-cone curve we would have had with other magnetic fields trace out curves that resemble the level-sets of the distribution function. $R=0$ corresponds exactly to the vertical line through the physical loss cone's vertex while $R=R_0$ corresponds to the physical loss cone. $R \in (0, R_0)$ covers the region between the vertical line and the physical loss cone in momentum-space. This $R$ identification successfully captures an important behavior of the model, namely that the plasma diffuses more quickly across $\theta$ than $x$.

\subsection{Deviations}
We analytically approximate $g_{\text{mod}}\left(R; R_0, \phi\right)$ by examining the level curves of the weighted simulation more closely. For $R\leq 0$, which captures the bulk of the distribution, we expect the distribution there to be close to Maxwellian. We set our prefactor for $R \leq 0$ to be unity. For $R \geq R_0$, which is the loss cone region, the distribution must vanish there. We set our prefactor for $R \geq R_0$ to be $0$. Note, this behavior has already been accounted for by the Heaviside theta function in the general form of $f_{\text{mod}}$ in Eqn. \eqref{eq:fmod}, but we still enforce this choice for $R$ for consistency. 

What remains is to find an analytic expression for $g_{\text{mod}} \left(R; R_0, \phi \right)$ in the region $R \in [0, R_0]$. To ensure smooth decay to the loss cone, we require that this function monotonically decreases and satisfies the following boundary conditions, 
\begin{equation}
g_{\text{mod}}\left(0; R_0, \phi\right) = 1 \qquad g_{\text{mod}} \left(R_0; R_0, \phi \right)=0.
\end{equation}

Naturally, we must recast the weighted simulation to be a function of the variable $R$. Then, $g_{\text{sim}}\left(R; R_0, \phi\right)$ corresponds to the average deviation of the numerical simulation from the truncated Maxwellian over the constant-$R$ curve. We find good agreement of this recasted $g_{\text{sim}}$ to a logarithmic function, which is shown in Fig. \ref{fig:curveanddev}. Our complete prefactor is the following, 
\begin{equation}
g_{\text{mod}} \left(R; R_0, \phi \right) = 
\begin{cases}
1 & R < 0 \\ 
1 - \log_{1+R_0}\left(1+R \right) & 0 \leq R < R_0 \\ 
0 & R \geq R_0
\end{cases}
\end{equation}
Overall, for a fixed $\phi$ value, we see that the $g_{\text{mod}}$ prefactor is a close match for most $R$ values except at low $R$. An example of $g_{\text{mod}}$ for fixed $\phi$ and $R_0$ is shown in Fig. \ref{fig:gmodunshiftandshift}.  

This low $R$ behavior stems from our assumption that all the constant-$R$ curves intersect the same vertex, which is not generally true as $R$ goes to zero. The assumption forces an infinite gradient at the vertex, which is un-physical. When $\phi$ is large, this is a safe assumption since the bulk of the distribution is Maxwellian and the numerical simulation's level curves appear compressed towards the vertex. However, when $\phi$ is small, the assumption worsens as the loss cone cuts into the bulk of the distribution. The numerical simulation's level curves then appear spread out from the vertex.

\subsection{Shifts}
\begin{figure}
\centering
\includegraphics[width=0.98\textwidth]{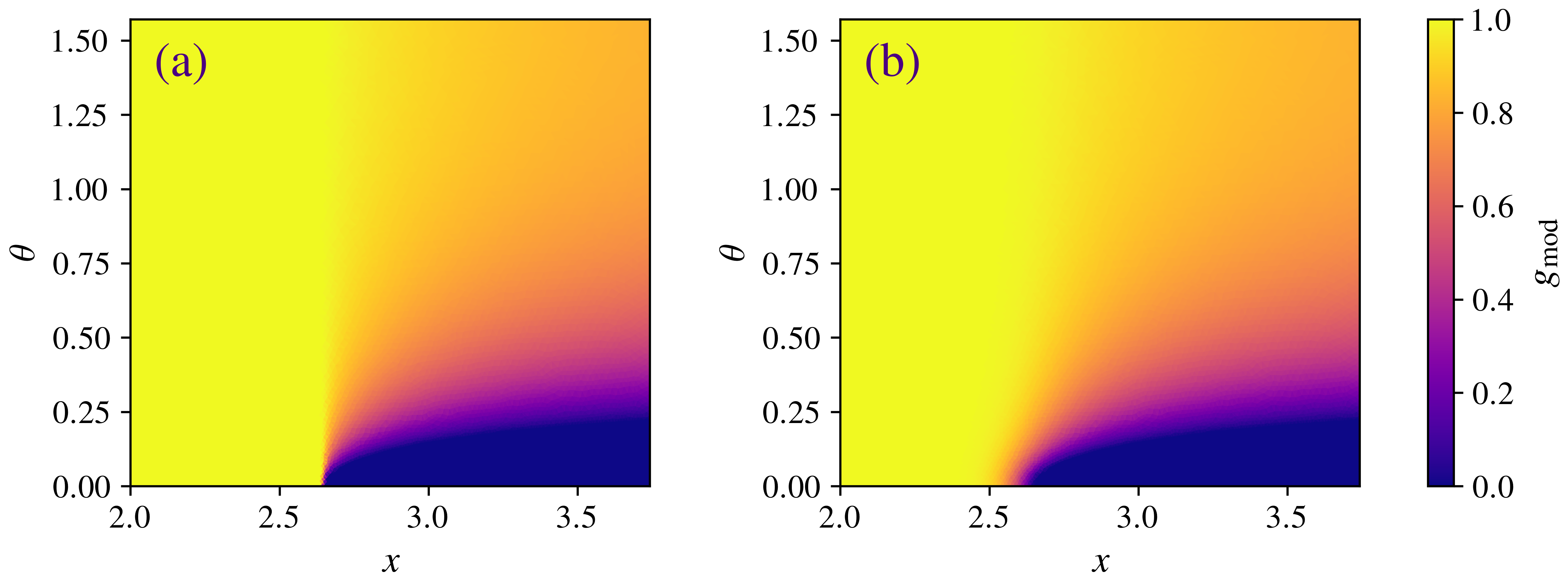}  
\caption{We plot $g_{\text{mod}}$ near the loss cone vertex for $\phi = 7$, $R_0 = 10$, and $Z_\perp = 1$. On the left, we use the unshifted parametrization which captures most of the smooth decay to the loss cone from $g_{\text{sim}}$ in Fig. \ref{fig:simandgsim}. On the right, we use the shifted parametrization, which noticeably improves the behavior at the vertex.}
\label{fig:gmodunshiftandshift}
\end{figure}

One way to correct the infinite gradient issue at the loss cone vertex is by shifting the constant-$R$ curves to lie at different vertices, which are the square roots of some corresponding effective potential $\phi_{\text{eff}}$.

The challenge is to find some simple smooth interpolation connecting $R$ and $\phi_{\text{eff}}$ that satisfies the following behaviors. The original loss cone must be preserved at $R=R_0$. As $R$ approaches $R_0$, the loss cones need to stay compressed, and $\phi_{\text{eff}} \approx \phi$. As $R$ approaches $0$, the loss cones need to spread out, and the shift is maximal. Lastly, as $\phi$ goes to infinity, we should also recover the unshifted prefactor. 

Let us consider the linear interpolation, where we choose to parametrize $R \in [0, R_0]$ and $\phi_{\text{eff}} \in [0, \phi]$ as the following, 
\begin{equation}
R(\alpha) = \alpha R_0 \qquad \phi_{\text{eff}} = \alpha \phi
\end{equation}
where $\alpha \in [0, 1]$. Note, the $R=0$ curve, which is the vertical line, is shifted to $x=0$. We solve for $R$, 
\begin{gather}
R(x, \theta) = \frac{x^2 R_0}{R_0 x^2 \sin^2\theta + \phi} .
\end{gather}
This interpolation lacks the correct limiting behavior as $\phi$ goes to infinity, but by recursively iterating $n$ times, we get the following interpolation,
\begin{equation}
R(x, \theta; n) = \frac{x^2 R(x, \theta; n-1)}{ R(x, \theta; n-1) x^2 \sin^2 \theta + \phi}.
\end{equation}
This recursion relation can be solved for any $n \in \mathbb{N}$,  
\begin{equation}
R_n(x, \theta) = \frac{R_0 \left(\phi - x^2\right)}{ \left(\phi/x^2\right)^n \left(\phi - x^2\right) + R_0 x^2 \sin^2\left(\theta\right) \left[\left(\phi/x^2\right)^n - 1\right]}.
\end{equation}

\begin{figure}
\centering
\includegraphics[width=0.98\textwidth]{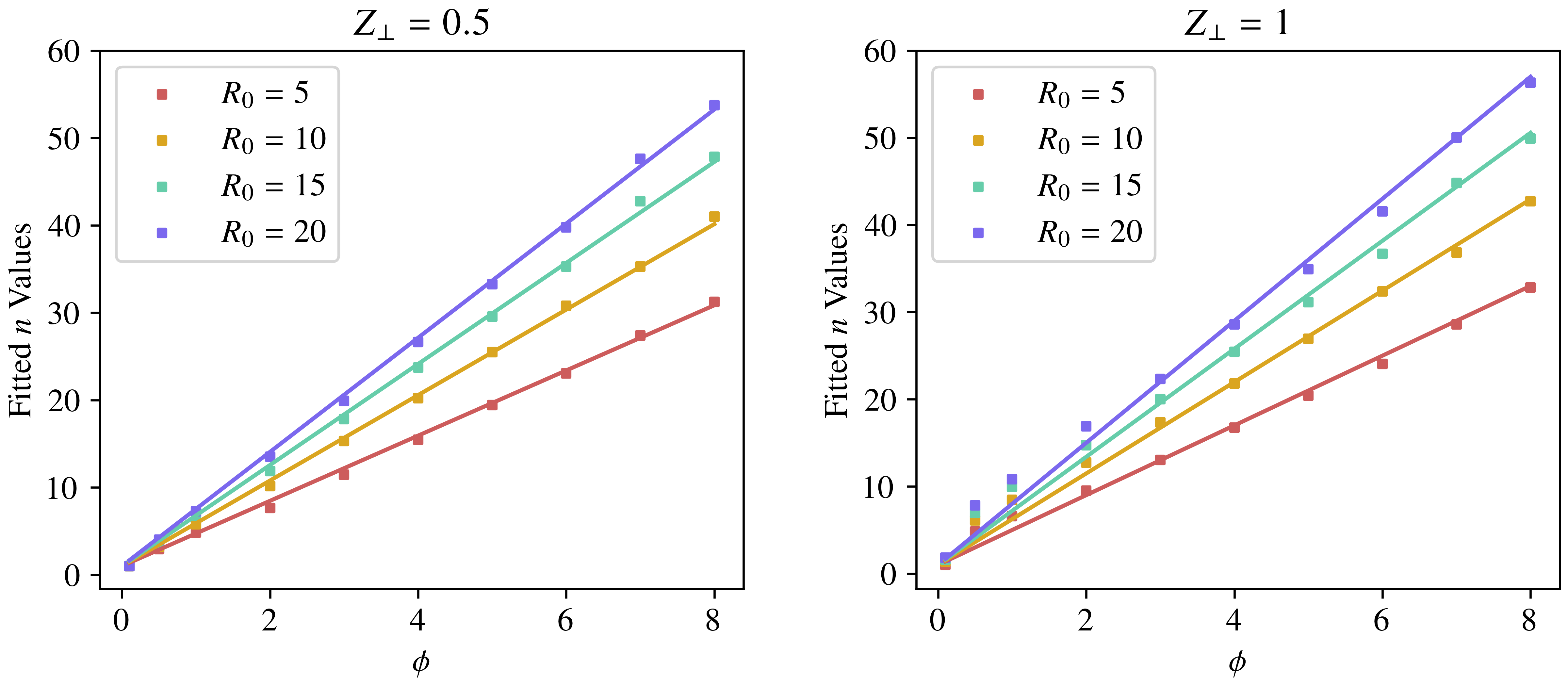}
\caption{Analytic fits for the simulated best-fit $n$ values for $Z_\perp = 0.5$ on the left and $Z_\perp = 1$ on the right. The fit is worse for $\phi < 1$, which makes sense since our model was found for $\phi > 3$. As $Z_{\perp}$ decreases, the loss cone curves should be shifted more due to decreased perpendicular diffusion. The $n$ values should then be correspondingly smaller, especially in the low $\phi$ region.}
\label{fig:nfits}
\end{figure}

For any $n \in \mathbb{N}$ and $(x, \theta)$ on the physical loss cone, $R_n(x, \theta) = R_0$, so the loss cone is preserved. This recursion interpolation also has the property that for a given $(x, \theta)$ that is not on the loss cone, $R(x, \theta, n-1) > R(x, \theta, n)$. Increasing $n$ compresses the constant-$R$ curves closer to the physical loss cone. To see this, as $n$ approaches $\infty$, we recover the unshifted model. Note, our prefactor case $g_{\text{mod}} = 1$ for unshifted $R<0$ is equivalent to setting $R=0$ for the region $x < \sqrt{\phi}$.  $R_{n\rightarrow \infty}(x, \theta)$ indeed approaches zero for that region and also approaches our unshifted $R$ in Eqn. \eqref{eq:unshiftedR} for $x > \sqrt{\phi}$. 

We can find analytic expressions for $n$ by fitting the prefactor from shifted $R$ to the weighted simulation for a range of $R_0$ and $\phi$. Specifically, for a given $R_0$ and $\phi$, we find $n$ value that minimizes an error metric described later in Eqn. \eqref{eqn:gof}. We require $n \geq 1$ since below one, the loss cone shapes for low $R$ curve towards $x = 0$ which is un-physical. In general, we find a nearly linear relationship between $n$ and $\phi$ and nearly square root dependence between $n$ and $R_0$, which we then fit to the simulated best-fit $n$ values. Below are two fits as seen in Fig. \ref{fig:nfits},
\begin{subequations} \label{eq:nfits}
\begin{gather}
n(R_0, \phi) = \left(\sqrt{1.57 R_0} + 0.93 \right) \phi + 1 \quad (Z_\perp = 0.5) \\
n(R_0, \phi) = \left(\sqrt{1.8 R_0} + 1\right)\phi + 1 \qquad (Z_\perp = 1). 
\end{gather}
\end{subequations}

We find that implementing some form of an interpolated shift improves the model in the low $\phi$ regime as seen in Fig. \ref{fig:gmodunshiftandshift}, but the precise form of $n(R_0, \phi)$ does not make much of a difference as long as $n$ monotonically increases with respect to $\phi$ and $R_0$ and $n(R_0, 0) > 1$.

We note here that our analytic model runs approximately 30 times faster than a single run of the finite-element solver, with both using a mix of C and Python. In conjunction with its independence of simulation codes, the analytic model has a substantial computational advantage over the numerical simulations.

\section{Error Metric Comparisons}\label{sec: gof}
\begin{figure}
    \centering
    \includegraphics[width=0.98\textwidth]{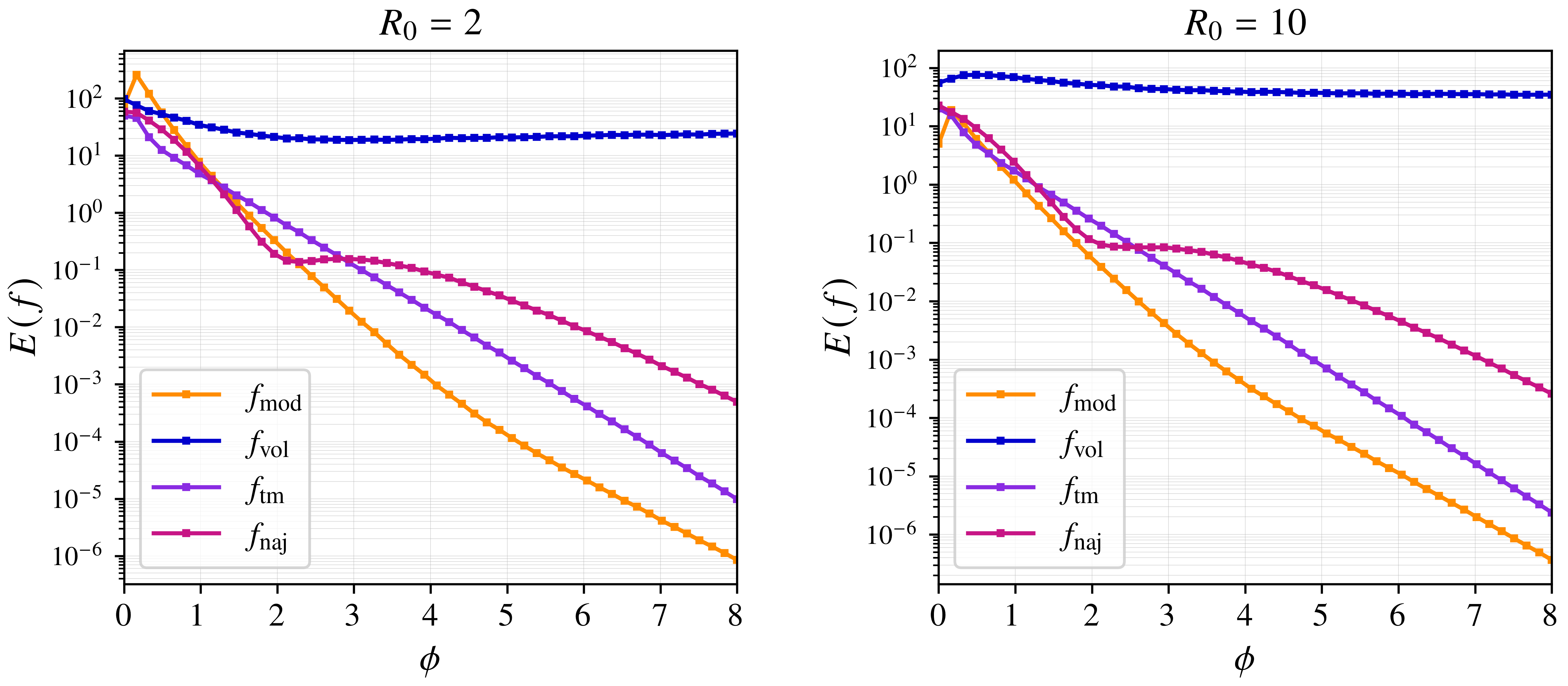}
    \caption{$E(f)$ of the proposed model, Volosov model, Najmabadi model, and truncated Maxwellian over $\phi \in [0, 8]$ for fixed mirror ratios. The spike at low $\phi$ where the proposed model has greater error than the truncated Maxwellian or Najmabadi model is much less pronounced at $R_0 = 10$ compared to $R_0 = 2$. For $\phi > 3$, the proposed model has less error by at least a factor of $10$ compared to the truncated Maxwellian.}
    \label{fig:error}
\end{figure}

\begin{figure}
    \centering
    \includegraphics[width=0.6\textwidth]{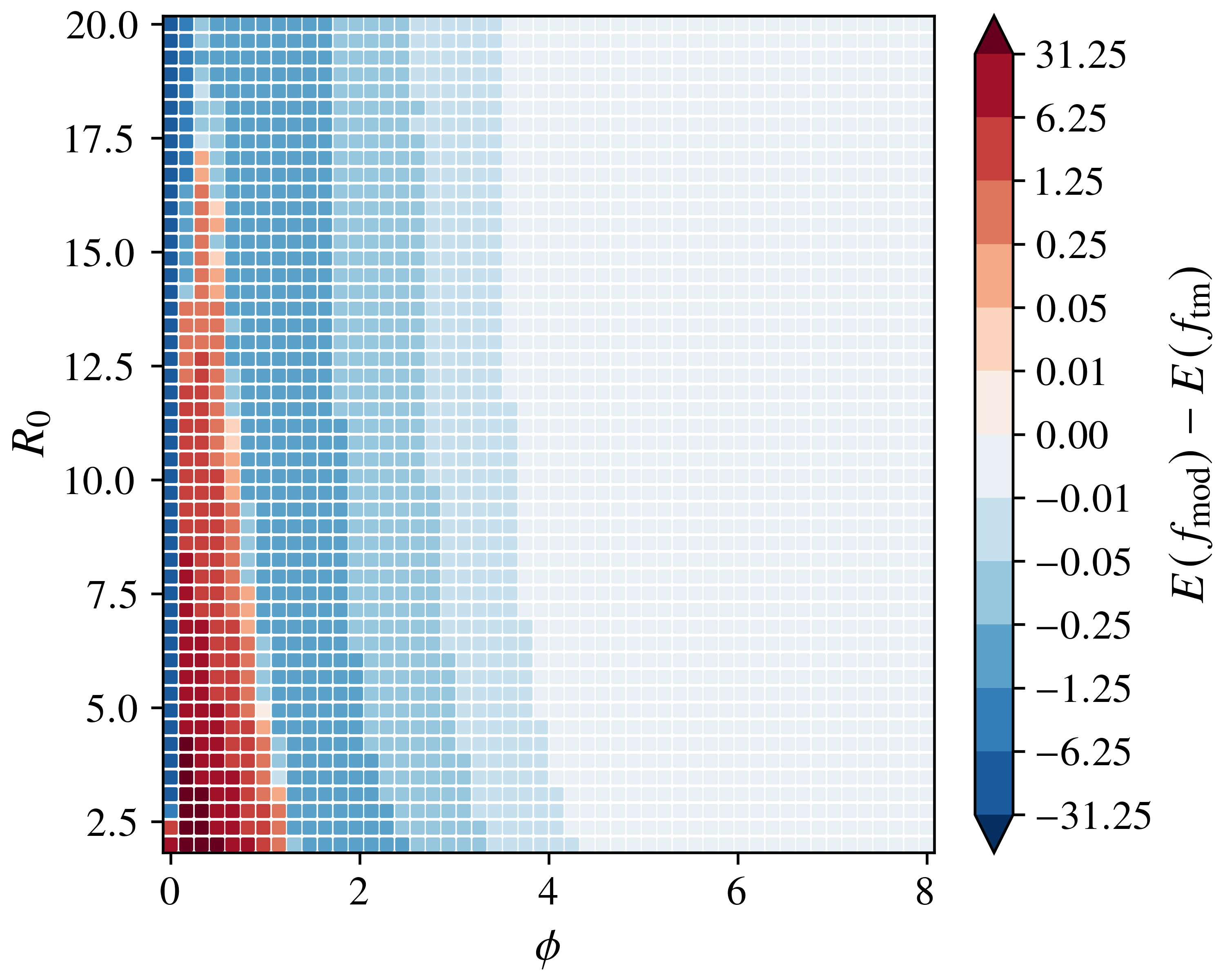}
    \caption{Difference of error metric results for the proposed model and the truncated Maxwellian, $E(f_{\text{mod}})-E(f_{\text{tm}})$. Blue regions where the difference is negative indicate where $f_{\text{mod}}$ outperforms $f_{\text{tm}}$ in fitting to the numerical simulations.}
    \label{fig:diff}
\end{figure}

\begin{figure}
    \centering
    \includegraphics[width=0.98\textwidth]{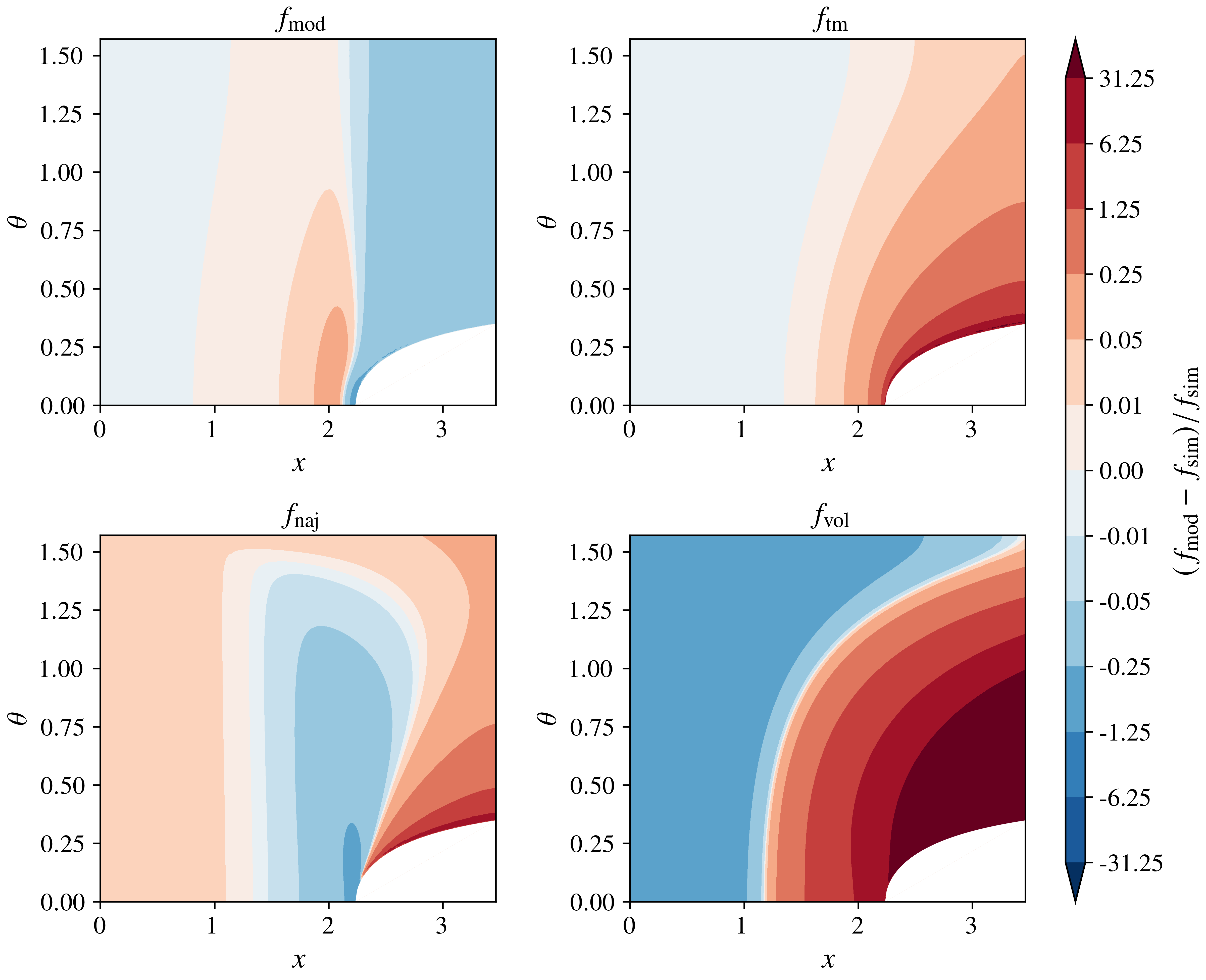}
    \caption{We compare the proposed model, truncated Maxwellian, Najmabadi model, and Volosov model side-by-side for $\phi=5$ and $R_0=5$ using relative error, $\left(f_{\text{mod}}-f_{\text{sim}}\right)/f_{\text{sim}}$. Red indicates regions where the analytic model overestimates the distribution; blue, underestimates.}
    \label{fig:comp}
\end{figure}

In this section, we demonstrate an error comparison of analytic models to the numerical simulation. We find that the proposed model $f_{mod}$ more closely matches the numerical simulation compared to the truncated Maxwellian model for most values of $\phi$ and $R_0$. The exception is a region spiking for low $R_0$ centered around $\phi = 0.2$. 

We used the following error metric, 
\begin{equation} \label{eqn:gof}
    \text{E(f)} = \sum_{(x, \theta)} \left(f(x, \theta; R_0, \phi) - f_{\text{sim}} (x, \theta; R_0, \phi)  \right)^2
\end{equation}
where $f$ is an arbitrary analytic model for the steady state particle distribution. All points in our momentum-space are weighted equally. Results for our proposed analytic, truncated Maxwellian, Najmabadi, and Volosov model for fixed $Z_\perp = 0.5$ are shown in Fig. \ref{fig:error}. We note that the proposed model $f_{\text{mod}}$ outperforms the truncated Maxwellian for regions of relatively good confinement, $R_0>5$ and $\phi > 2$, as shown in Fig. \ref{fig:diff}. This makes sense because our model was originally fitted to this region of stronger confining potential and larger mirror ratio. 

The Najmabadi model $f_{\text{naj}}$ does modestly better at fitting to the distribution than the truncated Maxwellian around $\phi \approx 2$, but does significantly worse in the limit of large $\phi$. This is surprising since $f_{\text{naj}}$ decays at the loss cone while the truncated Maxwellian does not. However, Najmabadi's model was developed for calculations of collisional end loss rates near the loss cone region. As the loss cone vertex moves further away from the bulk of the particles, fewer particles fall within the original intended region of Najmabadi's calculation. In Fig. \ref{fig:comp}, we see that $f_\text{naj}$ overestimates the distribution in the bulk and further along the loss cone boundary. We choose to look at the truncated Maxwellian in later analysis instead. Despite its inaccuracy at the loss cone, the truncated Maxwellian is a more consistent competitor to our model for much of parameter space. Surprisingly, the truncated Maxwellian can even outperform our proposed model in estimating the simulation at low $\phi$ and $R_0$.

\begin{figure}
    \centering 
    \includegraphics[width=0.98\textwidth]{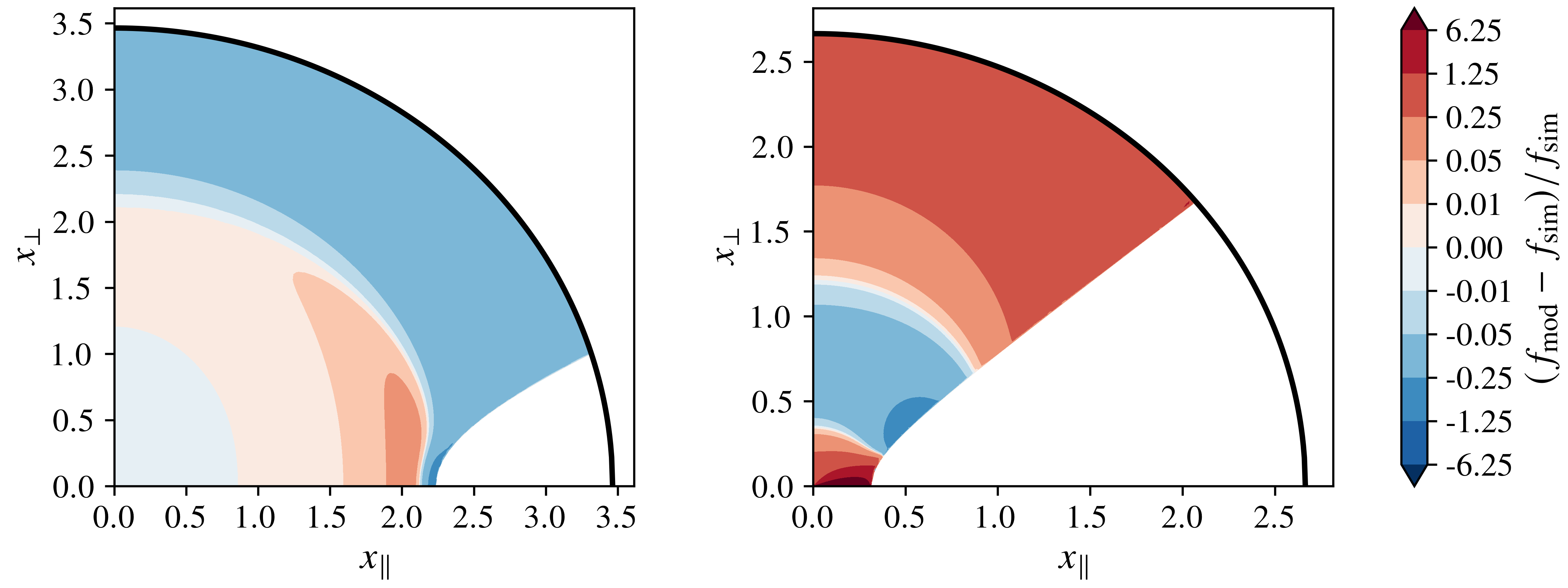} 
    \caption{Relative error of the proposed model compared to the simulation, $\left(f_{\text{mod}}-f_{\text{sim}}\right)/f_{\text{sim}}$. The left plot is for relatively good confinement $\phi = 5$ and $R_0 = 7$ while the right plot is for relatively worse confinement $\phi = 0.1$ and $R_0 = 2.5$, where the proposed model is known to be worse than other analytic models. The black line marks the simulation domain.}
    \label{fig:overunder}
\end{figure}

One possible reason for our model's worse fit at $\phi$ close to $0$ is due to the prefactor and normalization. The decay primarily impacts the tail, which is proportionally more of momentum-space when $\phi$ is small. The prefactor cuts out more of the distribution than in the simulation, leading to the normalization value to be non-trivially less than unity. However, since the value of the prefactor at low $R$ is fixed at one, the normalization will boost values at the bulk of the distribution while preserving the model's under-prediction at the tail. This is evident in Fig. \ref{fig:overunder}. This would also explain the narrow strip at $\phi \approx 0$ in $(\phi, R_0)$ parameter space where $f_{\text{mod}}$ far outperforms the truncated Maxwellian in approximating the steady state distribution. As $\phi$ approaches zero, the region of low $x$ between the origin and the vertex becomes negligible, and there are no values to be boosted. 

Choosing different $Z_{\perp} \in \{0.4, 0.6\}$ and greater source temperatures $T_s \in \{1, 2\}$ yields roughly the same results as our choice of $Z_\perp = 0.5$ and $T_s = 0.1$. Generally, as $T_s$ increases or as $Z_{\perp}$ decreases, the spiked region where our analytic model fails to outperform the truncated Maxwellian grows larger. Likewise, using the error metric that compares the weighted simulation and respective prefactors of the models led to a more definitive out-performance of our model relative to the truncated Maxwellian. The spike in $(\phi, R_0)$ space becomes negligible. This makes sense because we had originally found our prefactor by fitting to the weighted simulation.

Due to the significant error of the Volosov model, we have omitted analysis of that model from most of the following applications. This is evident by the much larger error in Fig. \ref{fig:error} and the significant overestimation of the distribution at the tail in Fig. \ref{fig:comp} due to the model's exponential dependence only on the perpendicular speed.

\section{Applications} \label{sec: apps}

\subsection{Stability Boundary}
In the limit of fast enough rotation, certain loss-cone instabilities such as the high-frequency convective loss-cone (HFCLC), drift cyclotron loss-cone (DCLC), and Dory-Guest-Harris (DGH) mode can be stabilized. We are interested in the stability threshold, or minimum value of confining potential $\phi$ for a given mirror ratio $R_0$ that suppresses a given loss-cone mode. In this paper, we restrict ourselves to studying the HFCLC mode, comparing the predictions for the stability boundary from $f_{\text{mod}}$ to other analytic models and numerical simulations. For more context on stability threshold problems, see \citep{kolmes_loss-cone_2024, volosov_characteristics_1969, turikov_effect_1973}. 

We first review the stability condition for the HFCLC mode. We define the perpendicular projected distribution as
\begin{equation}
    \psi_a(x_\perp^2) \coloneqq \bar{\psi} \int_{-\infty}^{\infty} f_a(x_\perp^2, x_\parallel^2) d x_\parallel
\end{equation}
where $\bar{\psi}$ is the normalization constant. Note that we use the parallel and perpendicular components of momentum, not $x$ and $\theta$. Perpendicular monotonicity is a sufficient condition for stabilizing the HFCLC mode, 
\begin{equation}
    \frac{\partial \psi}{\partial z} \leq 0 \qquad (\forall z\geq0)
\end{equation}
where $z \coloneqq x_\perp^2$. The partial derivative with respect to $z$ for both the truncated Maxwellian and Volosov model can be calculated analytically as done in \citet{kolmes_loss-cone_2024}. However, due to the parametrization required for our analytic model, the partial derivative with respect to $z$ for $f_{\text{mod}}$ must be calculated numerically. 

\begin{figure}
    \centering
    \includegraphics[width=0.6\textwidth]{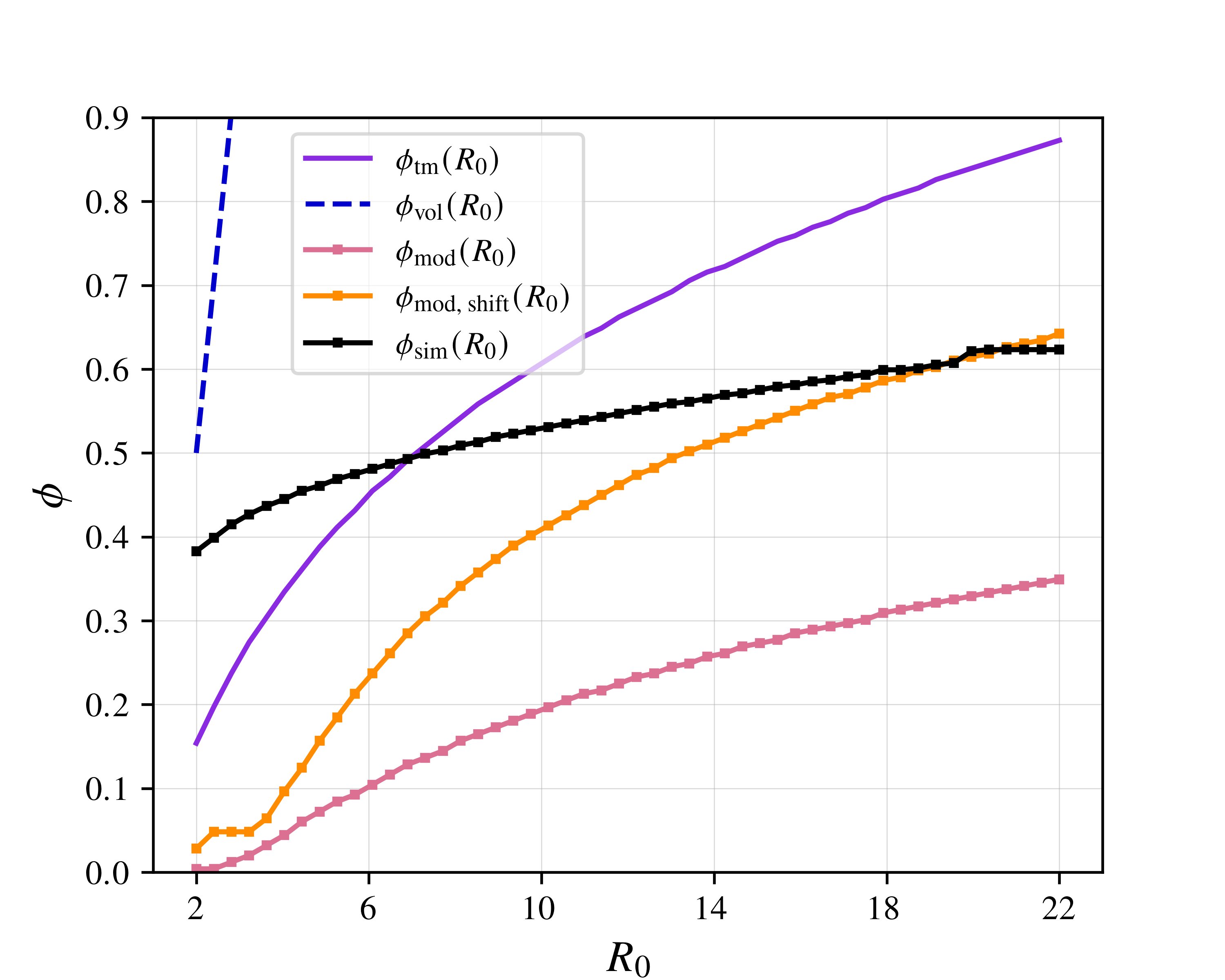} 
    \caption{Stability boundary curves for analytic distributions and the numerical simulations.}
    \label{fig:stabilitycurves}
\end{figure}

We find the stability boundaries $\phi(R_0)$ of four distributions compared to the numerical simulations in Fig. \ref{fig:stabilitycurves}. The stability boundary of the Volosov model can be shown to be directly proportional to $R_0$, and it does not match our simulation's results. The truncated Maxwellian over-predicts the stability values for $R_0 < 7$ and under-predicts the stability values elsewhere. As for our proposed model, we get a closer fit to the numerical simulation's stability boundary when we use the shifted as opposed to the unshifted parametrization. Both of our proposed distributions under-predict the simulation stability value, but shifting lifts the stability values of the model with unshifted parametrization. 

This can be explained by understanding the region in $(\phi, R_0)$ space where the truncated Maxwellian outperforms $f_{\text{mod}}$ in our error analysis. The stability values are precisely at low $\phi < 1$ values, where our model over-predicts at low $x$ and under-predicts at higher $x$. This tendency of over and under estimation is visible in Fig. \ref{fig:overunder}, which is with respect to $x_\parallel-x_\perp$ coordinates. The cumulative effect is to boost the projected distribution at low $z$, enough to satisfy the perpendicular monotonicity condition when the numerical simulation does not. Increasing the mirror ratio $R_0$ improves the performance of the model compared to the truncated Maxwellian, which is what we observe in the stability boundary curves. 

\subsection{Fusion Yield}\label{sec: fusionyields}
One use of analytic distribution functions is more quickly calculating fusion yields. Since fusion yield is sensitive to the tail of the distribution, the accuracy of a model in capturing behavior near the loss cone and further away from the bulk of the distribution matters more. We compare the fusion yield of the deuterium-deuterium reaction from the numerical simulation to various analytic models. We find that proposed model more closely estimates the fusion yield compared to the truncated Maxwellian. However, the model consistently underestimates fusion yield as a result of over-suppressing values at the loss cone. For further context on fusion yield problems, see the following \citep{kalra_fusion_1988, nath_thermonuclear_2013, kolmes_fusion_2021, Xie2023, Kong2024, fetsch2025arxiv}.

To find fusion yield, the fusion reaction rate $Y$ between two ion species can be expressed as the following integral expression, 
\begin{equation}
    Y(f_a, f_b) = \int d^{3} \textbf{v}_a d^3 \textbf{v}_b \sigma(w) w f_a(\textbf{v}_a) f_b(\textbf{v}_b)
\end{equation}
where $w \coloneqq |\textbf{v}_a - \textbf{v}_b|$ is relative speed, $f_a$ and $f_b$ are the distribution functions of the ion species, and $\sigma$ is the respective cross section between species $a$ and $b$. This six-dimensional integral can be reduced to a five-dimensional integral by using cylindrical symmetry for one of the ion species. To numerically evaluate the fusion yield integral, we use the VEGAS Monte Carlo algorithm as described in \citet{lepage_new_1978}. 

\begin{figure}
    \centering 
    \includegraphics[width=0.98\textwidth]{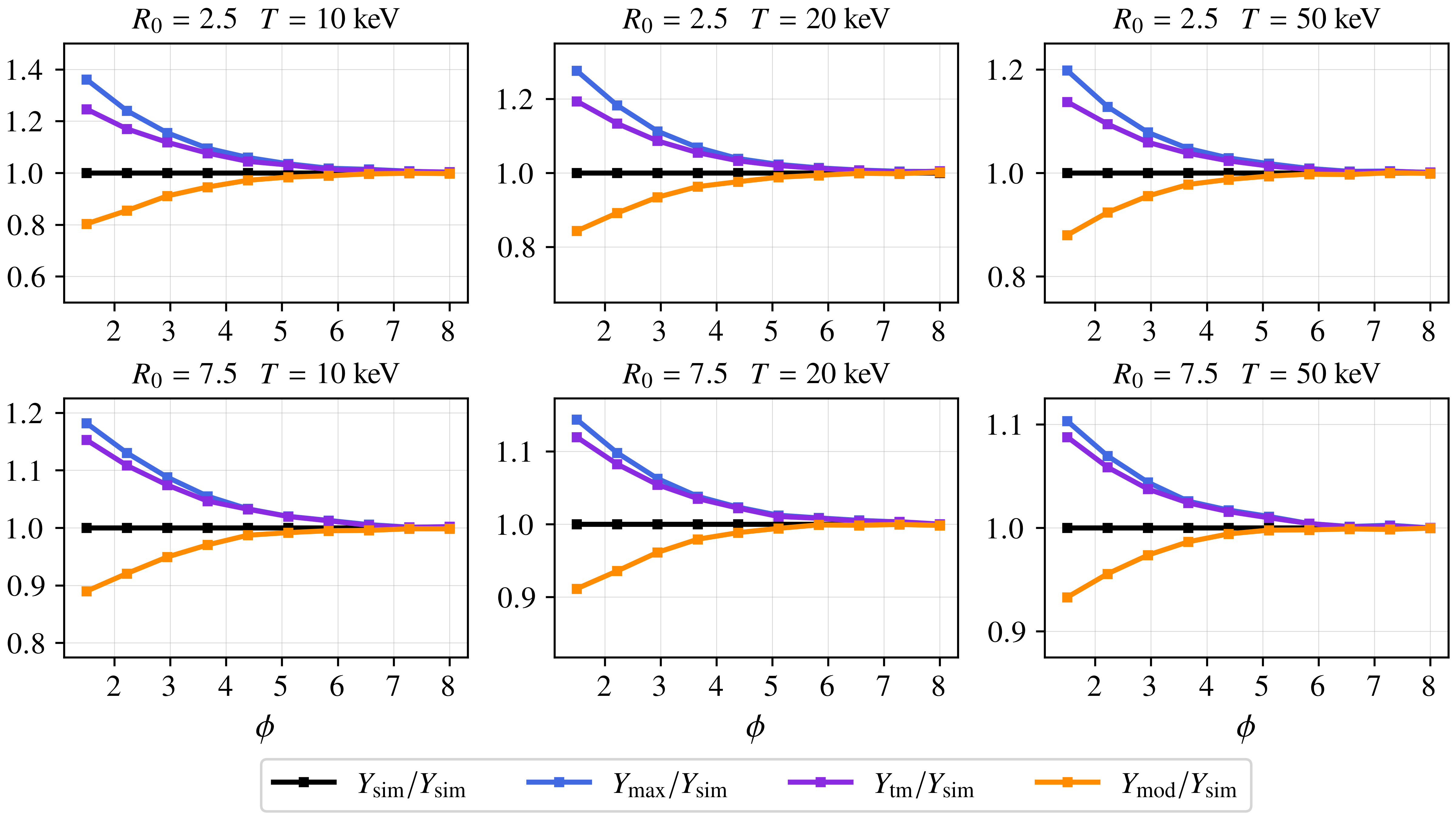} 
    \caption{We plot relative fusion yields $Y_f/Y_{\text{sim}}$ for fixed temperatures and mirror ratios. We vary confining potentials $\phi \in [1, 8]$. Note, as $\phi$ increases, all the models converge to the simulation results as they all look essentially Maxwellian in this limit.}
    \label{fig:phiyields}
\end{figure}

\begin{figure}
    \centering 
    \includegraphics[width=0.98\textwidth]{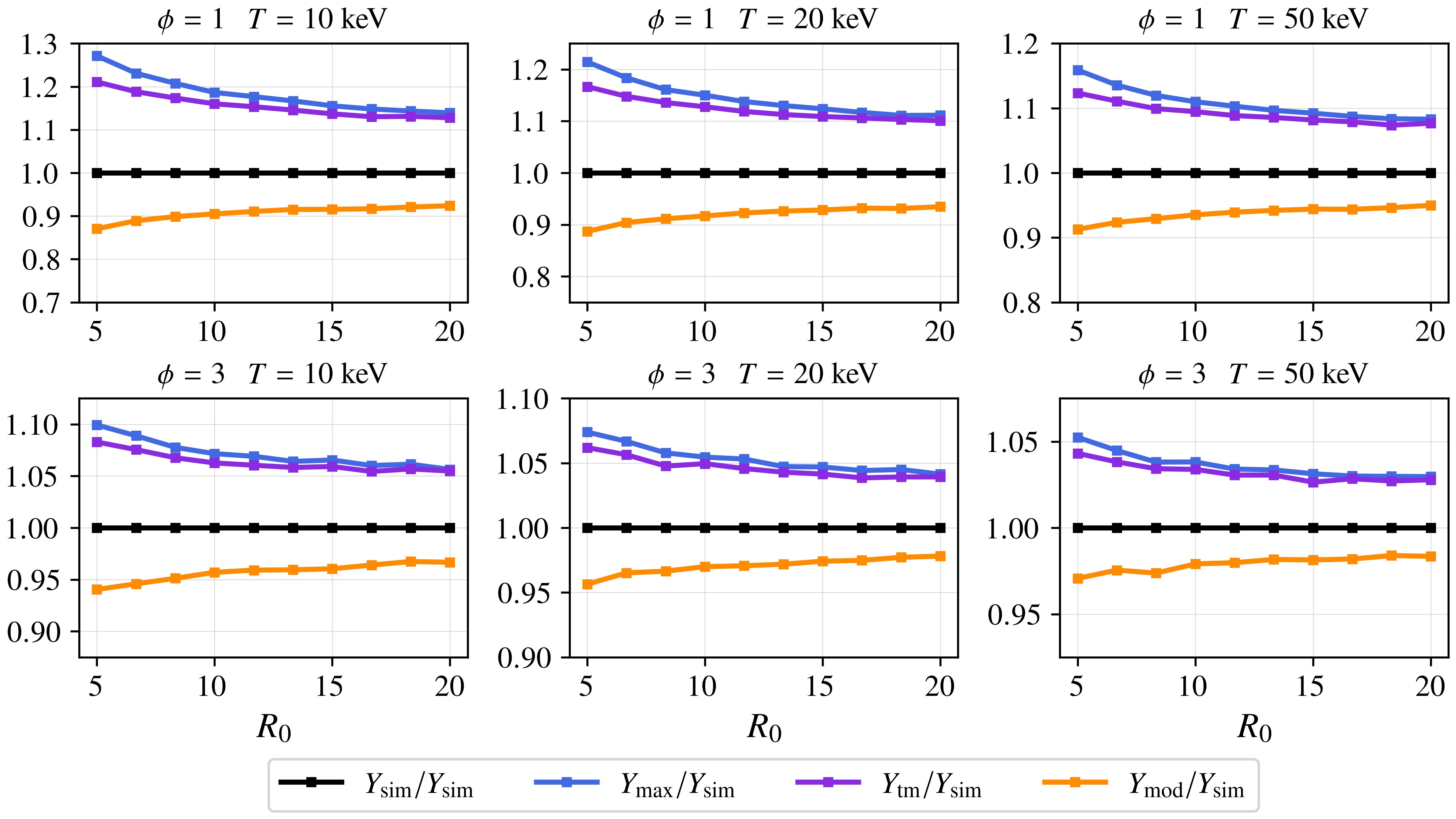} 
    \caption{We plot relative fusion yields $Y_f/Y_{\text{sim}}$ for fixed temperatures and confining potentials. We vary mirror ratios $R_0 \in [5, 20]$. The model better predicts fusion yield compared to the Maxwellian and truncated Maxwellian for all mirror ratios.}
    \label{fig:r0yields}
\end{figure}

For simplicity, we find the reaction rate of a single ion species, deuterium, for the deuterium-deuterium reaction at temperatures $10$, $20$, and $50$ keV for different mirror ratios and confining potentials. The ion species then has $Z_\perp$ value of $0.5$ exactly. We find the fusion yields of three models, the Maxwellian, truncated Maxwellian, and our proposed model as well as the fusion yield of the numerical steady-state simulation. Our yield calculations are shown in Fig. \ref{fig:phiyields} for varying confining potential, and Fig. \ref{fig:r0yields} for varying mirror ratios.

When ranging over both confining potentials and mirror ratios, we find that the proposed model has less relative error compared to the truncated Maxwellian. We first note that the model under-predicts the fusion yield while the truncated Maxwellian over-predicts. The reason can be found in the tail behavior. The Maxwellian slightly over-predicts compared to the truncated Maxwellian, with the only difference between the distributions being the removal of the loss cone particles and resultant change in normalization constant. From Fig. \ref{fig:overunder}, since we are considering relatively good confinement with $\phi > 1$ and $R_0 \geq 2$, it is evident that our model over-suppresses the distribution values at the tail, leading to a consistent under-prediction of fusion yield.

\subsection{Free Energy}
Finally, we consider the calculation of free energy associated with the steady-state particle distribution. Free (or available) energy is the maximum amount of energy extractable from a given initial distribution after specifying rules on allowed phase-space rearrangements. Equivalently, free energy also provides a bound on the amount of energy that can go into instabilities before reaching a distribution called the ground state, which is necessarily stable. If a relation can be found between available energy and another stability-related quantity, available energy could serve as a computationally simpler proxy measurement for that other quantity, as demonstrated for turbulent energy flux in tokamaks and stellarators \citep{mackenbach_available_2022, mackenbach_available_2023, mackenbach_available_2023-1}. We study the free energy of the proposed model, truncated Maxwellian, and numerical simulation subject to two different sets of rearrangement constraints.

We briefly review necessary background on free energy and relevant constraints. Free energy subject to the sole constraint that all rearrangements must preserve phase-space volumes is called the Gardner free energy \citep{gardner_bound_1963}. Reaching a ground state is equivalent to rearranging the initial velocity-space distribution into a distribution $f$ that is monotonically decreasing with respect to energy. The energy of a distribution is commonly chosen as the total kinetic energy $W$. The available energy is then calculated as the difference of the initial state's kinetic energy and the ground state's kinetic energy. However, in constructing a coarse-grained model of the distribution function that averages over some scale, we often perceive diffusion, or phase mixing. Free energy subject to the constraint that rearrangements average the densities of phase-space volumes is called diffusive free energy~\citep{fisch_free_1993}. Although diffusive free energy is more appropriate for the collisional diffusion behind our model, it is sufficient to calculate the Gardner free energy since it has been shown that the two free energies are arbitrarily close in the continuous limit \citep{kolmes_recovering_2020}. When we refer to the unconstrained free energy, we are referring to the Gardner free energy. 

\begin{figure}
    \centering 
    \includegraphics[width=0.98\textwidth]{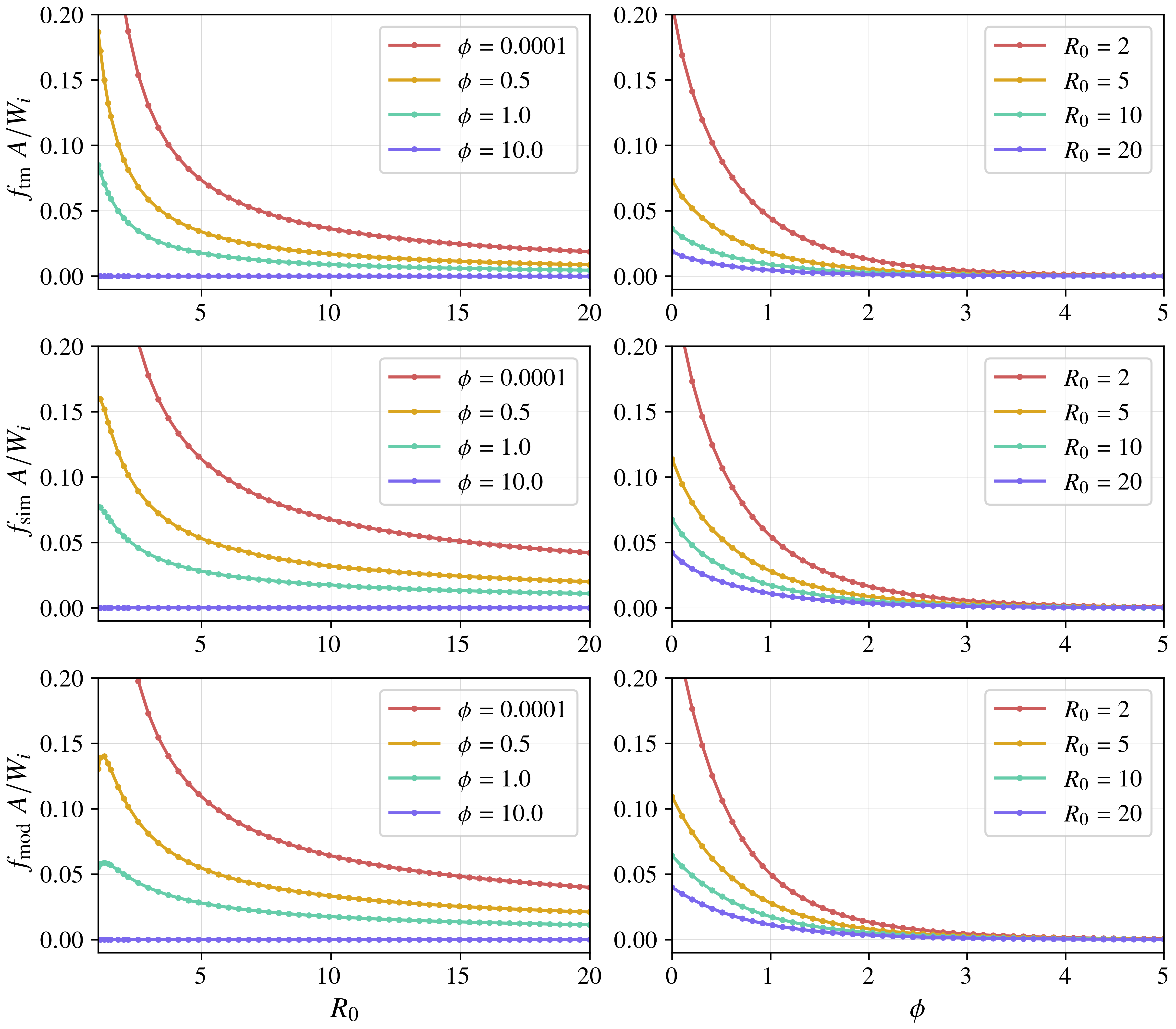} 
    \caption{Accessible energy fraction $A/W_i$ for the unconstrained case.}
    \label{fig:unconstrained}
\end{figure}

\begin{figure}
    \centering
    \includegraphics[width=0.98\textwidth]{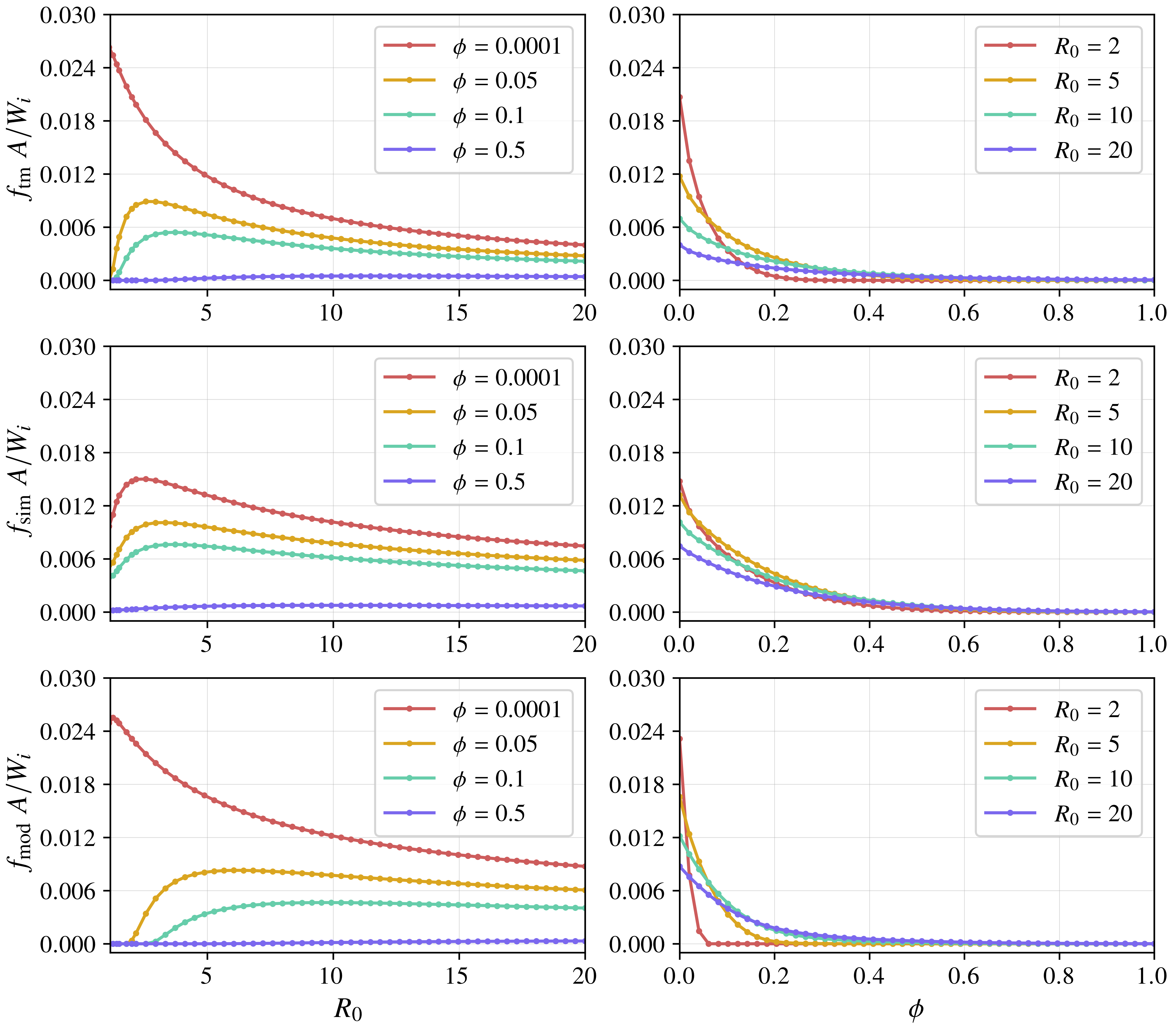} 
    \caption{Accessible energy fraction $A/W_i$ for the constrained case.}
    \label{fig:constrained}
\end{figure}

Conservation of phase-space volume is often not a sufficient description of the allowed phase-space rearrangements of the plasma for any arbitrary mode. It is possible to impose additional constraints on allowed rearrangements such as the conservation of adiabatic invariants \citep{helander_available_2017,helander_available_2020}. An important class of loss-cone instabilities are flute-like modes, such as the HFCLC mode, whose wave numbers $k$ vanish in the direction of the magnetic field. One additional constraint for these modes is that only the projected distributions corresponding to a specific $x_\perp$ value,
\begin{equation}
    f_P(x_\perp) = \int_{-\infty}^{\infty} f(x_\parallel, x_\perp) d x_\parallel, 
\end{equation}
can be swapped with one another \citep{kolmes_loss-cone_2024}. This comes from the fact that the quasilinear diffusion operator cannot differentiate between phase-space elements of the same $x_\parallel$. When we refer to constrained free energy, we are referring to the available energy subject to both constraints of phase-space volume conservation and this flute-like loss cone constraint. For a more extensive discussion of calculating available energy, see \cite{kolmes_available_2020, helander_available_2017, helander_available_2020}.

To compare the free energy of different initial distributions, we are interested in the accessible energy fraction, or the ratio of free energy $A$ to the total initial energy $W_i$. For our numerical calculations, we simulate more of velocity-space by setting $K=12$. We also choose a smaller source temperature $T_s = 0.01$ so that the source does not disturb the steady-state distribution for almost negligible confining potentials.

For unconstrained free energy, the left column of Fig \ref{fig:unconstrained} shows the accessible energy fraction's dependency on mirror ratio $R_0$ for initial distributions for fixed confining potentials.  When $\phi=10$, all the distributions have zero available energy, which is what we expect. For large $\phi$, the loss cone moves further away from the bulk of the distribution, and thus, the initial distribution approaches the Maxwellian, which is monotonically decreasing in energy. $f_{\text{mod}}$ closely matches the numerical simulations for $R_0>2.5$, even for low $\phi$ values, and outperforms the truncated Maxwellian in the $A/W_i$ calculations. However, for low $R_0$ values, $f_{\text{mod}}$ underestimates $A/W_i$ and exhibits a non-monotonic behavior close to $R=1$. Our model was optimized for the $R_0>5$ regime, so it makes sense for it to do poorly at low $R_0$. 

The right column of Fig. \ref{fig:unconstrained} shows the accessible energy fraction's dependency on confining potential $\phi$ for different initial distributions for fixed mirror ratios. The $f_{\text{mod}}$ curves seem to closely match the numerical simulation $A/W_i$ curves for all $R_0$ and $\phi$. Meanwhile, the truncated Maxwellian underestimates $A/W_i$ in the low $\phi$ limit and decays faster than the numerical simulation.

For constrained free energy, the left column of Fig. \ref{fig:constrained} shows the accessible energy fraction's dependency on mirror ratio $R_0$ for initial distributions with the additional flute-like mode constraint. The chosen fixed confining potentials are smaller than the unconstrained counterparts. In the low $\phi$ limit, both the truncated Maxwellian and $f_{\text{mod}}$ find zero available energy. In cutting out values near the loss cone, the prefactor of $f_{\text{mod}}$ boosts values for $x_\parallel \approx 0$ and low $x_\perp$, which can force the initial projected distribution to be monotonically decreasing with respect to $x_\perp$. The truncated Maxwellian gets closer to the immediate shape of the numerical simulation's accessible energy fraction curves. In the high $\phi$ limit, the calculated available energy of $f_{\text{mod}}$ approaches the numerical simulation's available energy values while the truncated Maxwellian underestimates. 

The right column of Fig. \ref{fig:constrained} shows the accessible energy fraction's dependency on confining potential $\phi$ for different initial distributions with the flute-like mode constraint. The chosen fixed mirror ratios are the same as the unconstrained case. $f_{mod}$ appears to overestimate $A/W_i$ and then decay more quickly than the numerical simulation while the truncated Maxwellian underestimates $A/W_i$ (except for $R_0=2$) while decaying closer to the rate that the numerical simulation does.

\section{Discussion}\label{sec: final}
In this paper, we have presented a closed-form analytic model of the steady state particle distribution function for low collisionality mirrors traps like tandem or centrifugal traps subject to arbitrary confining potentials. The proposed distribution has the following desired qualities: 1) recovering the Maxwellian in the limit that the confining potential goes to infinity, 2) smoothly decaying to zero at the loss cone boundary, 3) diffusing more rapidly in pitch angle than speed, and 4) maintaining cylindrical symmetry. Motivated by the level curves similarity to loss-cone shapes, we formulate both an ``unshifted'' and ``shifted'' parametrization for the model. Although the shifted formulation requires an extra parameter depending on the $Z_\perp$ value, a simple approximation suffices, and we provide two such fits. Otherwise, our analytic model is written explicitly below and remains flexible enough to be modified for different uses.

\begin{equation}
    f_{\text{mod}}(x, \theta; R_0, \phi) = A \left[g_{\text{mod}}(x, \theta; R_0, \phi) \left(\pi^{-3/2} e^{-x^2}\right) \right]\Theta\left(\phi + R_0 x^2 \sin^2(\theta) - x^2 \right)
\end{equation}
\begin{equation}
g_{\text{mod}} \left(R; R_0, \phi \right) = 
\begin{cases}
1 & R < 0 \\ 
1 - \log_{1+R_0}\left(1+R \right) & 0 \leq R < R_0 \\ 
0 & R \geq R_0
\end{cases}
\end{equation}
\begin{equation}
R(x, \theta) = \frac{x^2 - \phi}{x^2 \sin^2(\theta)} \quad \text{(unshifted)} 
\end{equation}
\begin{equation}
R_n(x, \theta) = \frac{R_0 \left(\phi - x^2\right)}{ \left(\phi/x^2\right)^n \left(\phi - x^2\right) + R_0 x^2 \sin^2\left(\theta\right) \left[\left(\phi/x^2\right)^n - 1\right]} \quad \text{(shifted)}
\end{equation}
where fits for $n$ are provided for $Z_\perp = 0.5$ and $Z_\perp = 1$ in Eqn. \eqref{eq:nfits}. 

We then compared several analytic models to the numerical steady-state simulated distribution with an error metric. We find that our model has nearly a factor of $10$ less of error compared to the truncated Maxwellian for mirror ratio $R_0 > 5$ and confining potential $\phi > 2$. In this region of relatively good confinement, our model tends to underestimate the distribution values for larger values of momentum, more severely at the loss cone vertex. In general, our model does better in approximating the distribution than the truncated Maxwellian except for the spike at $\phi = 0.2$ in $(R_0, \phi)$ space. We have also compared the truncated Maxwellian to other models and found that it performs better than expected despite un-physical behavior at the loss cone. The truncated Maxwellian is a simpler alternative to our model that has the trade-off of somewhat worse performance for much of parameter space. Note, this model is better suited for bulk-behavior applications. The truncated Maxwellian is consistently worse than our model for tail-specific applications such as fusion yield. 

These analytic functions are often useful heuristic devices for studying the stability of loss cone modes and tail behaviors. We have explored three applications for our model, the HFCLC stability boundary calculation, fusion yield for different temperatures of a D-D plasma, and both unconstrained and constrained (by flute-like loss cone mode) available energy. There are a few key trends, which we review here. For the low $\phi \ll 1$ regime, our model boosts the distribution values at low momentum, which causes a noticeable under-prediction of the minimum stability value and available energy in the constrained case. For the region of relatively good confinement, our model over-suppresses the distribution values at high momentum, in particular at the loss cone boundary. Specifically, this behavior results in under-prediction of fusion yields, which are sensitive to the higher energy particles at the tail. Note, for relatively good confinement or sufficiently high mirror ratios for small confining potentials, our model will often more closely recover the limiting behavior of the numerical simulation compared to the truncated Maxwellian model.

It can be anticipated that improvements to our model functions can be made by refining the shift. The shift was based off a recursive linear interpolation, but other interpolations (polynomial, logarithmic, etc.) between $R$ and $\phi_\text{eff}$ are options. By either using another interpolation or finding a closed form expression for the $n$ parameter that accounts for different $Z_{\perp}$ values, a further simplified expression for the model could be found that does not compromise the fit of the model to the numerical simulations. 

Our analytic distribution can also be used to calculate radiation loss quantities in rotating mirrors to judge the viability of the configuration \citep{mlodik_sensitivity_2023,munirov_bremsstrahlung_2023,ochs_tailsuppression_2024}. Additionally, the yield calculation shown in Section \ref{sec: fusionyields} could be extended to calculate the birth distribution of fusion products in phase space, which is important for alpha-extraction problems \citep{gudinetsky_removal_2025,fisch_interaction_1992,fisch_alphamirror_2006,zhmoginov_alpha_2008,mynick_transport_1994, bierwage_energy-selective_2022}. Comparisons between the proposed model and numerical simulations could then be made with respect to these calculations to judge the accuracy of our model. Improving upon these heuristic models is essential to better understanding and more quickly calculating quantities of interest in centrifugal and tandem mirrors.

\section*{Acknowledgments}
The authors would like to thank Maxwell Rosen, Tal Rubin, and Alexander Glasser for helpful conversations. 

\section*{Funding}
This research was performed with support from the Program in Plasma Science and Technology (PPST), under US DOE contract number DE-AC02-09CH11466. G.L. was supported by Princeton University's Office of Undergraduate Research Undergraduate Fund for Academic Conferences through the Hewlett Foundation Fund. This work was also supported by APRA-E Grant No. DE-AR0001554.

\section*{Declaration of Interests}
The authors report no conflict of interest. 

\section*{Author ORCID}
G. X. Li, https://orcid.org/0009-0000-4295-8012; E. J. Kolmes, https://orcid.org/0000-0001-5303-5299; I. E. Ochs, https://orcid.org/0000-0002-6002-9169; N. J. Fisch, https://orcid.org/0000-0002-0301-7380.

\bibliographystyle{jpp}

\bibliography{master}

\end{document}